\documentclass[aps, pre, floatfix]{revtex4}
\usepackage{epsfig} 

\def\la{\langle}
\def\ra{\rangle}
\begin{document}

\title{Rotational correlation and dynamic heterogeneity\\ in a kinetically constrained lattice gas}

\author{Albert C. Pan}

\affiliation{Department of Chemistry, University of California,
Berkeley, CA 94720-1460}

\date{\today}

\begin{abstract}
We study dynamical heterogeneity and glassy dynamics in a kinetically
constrained lattice gas  model which has both translational and rotational degrees
of freedom.  We find that the rotational diffusion constant
tracks the structural relaxation time as density is increased whereas 
the translational diffusion constant exhibits a strong decoupling.  
We investigate distributions of exchange and
persistence times for both the rotational and translational degrees of
freedom and compare our results on the distributions of rotational exchange
times to recent single molecule studies.

\end{abstract}


\maketitle

\section{Introduction}

Glassformers are dynamically heterogeneous.  Neighboring regions,
nanometers in size, can have local relaxation times which differ by
several orders of magnitude \cite{EdigerReview, VandenboutScience}.
  Experimental measures of heterogeneous dynamics often 
probe rotational degrees
of freedom.  For example, Deschenes and Vandenbout
\cite{VandenboutScience, Vandenbout2001} have measured the rotation of single probe
molecules
in a polymer film near the glass transition. 
Here, we investigate a kinetically constrained lattice gas model
with translational and rotational degrees of freedom which captures many of the 
essential features seen in these experiments such as stretched
exponential
decay of rotational autocorrelation functions as temperature is
decreased and heterogeneous distributions of exchange times. 

In section \ref{Models}, we present the model and the computational
methods
used.  Section \ref{hetero} demonstrates the existence of
heterogeneous
dynamics.  We also present ensemble measurements which
display
a precipitous dynamical slowdown and breakdown of mean field
dynamical relations.  Section
\ref{dist}
measures distributions of persistence and exchange times.  Finally,
section \ref{exp} compares these distributions to those observed
recently in single molecule studies.

\begin{figure*}
\centering
\includegraphics*[width = 6in]{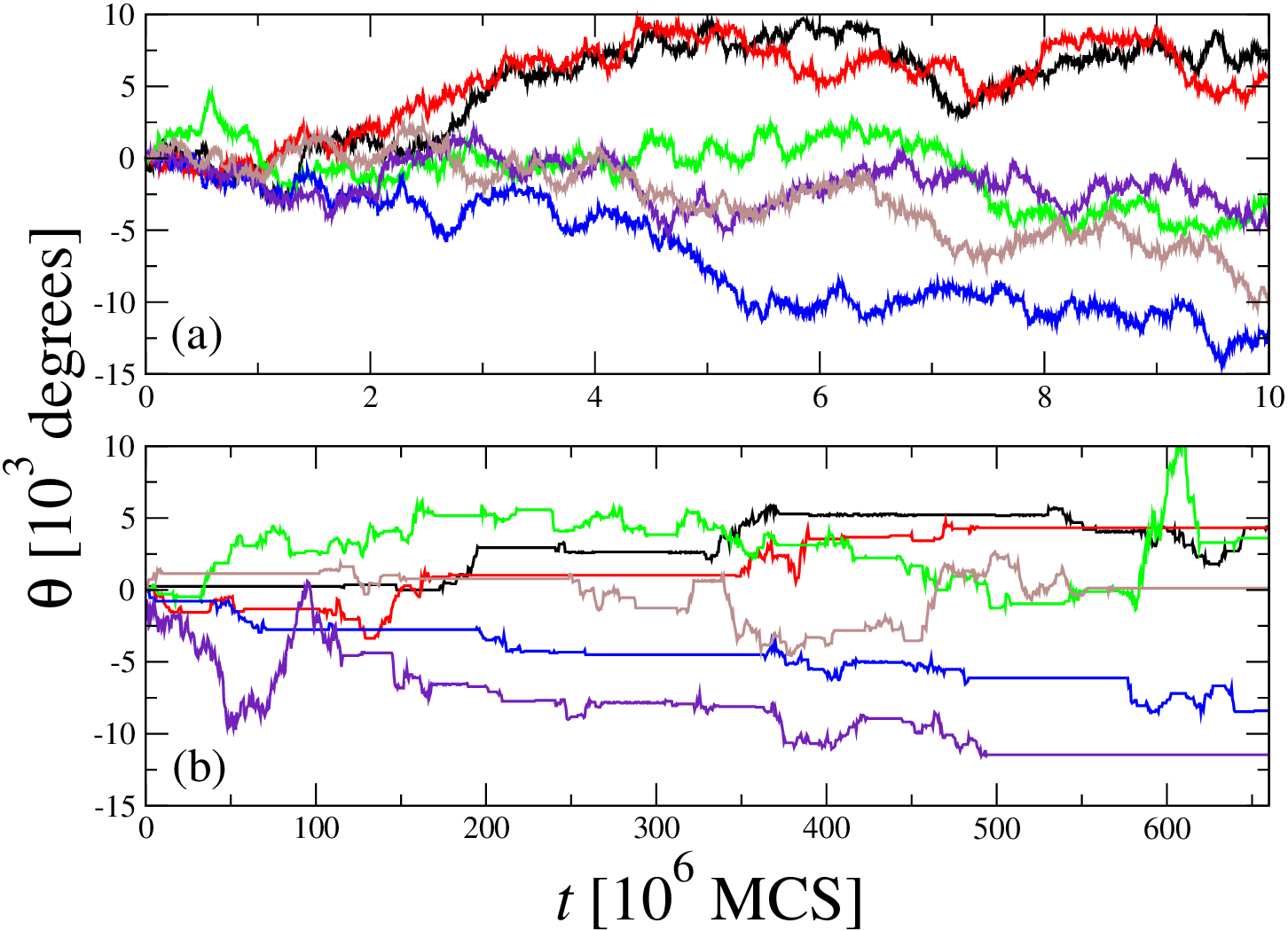}
\caption{Rotational trajectories of single particles at (a) low density ($\rho = 0.30$) and (b)
  high density ($\rho = 0.77$).}
\label{probes}
\end{figure*}

\section{Models and Computational Details}
\label{Models}

One route to understanding dynamical heterogeneity relies on the
presence
of local steric constraints on the movement of particles which make
themselves felt to an increasing degree as temperature is lowered 
(or density is increased).  
The kinetically
constrained
lattice gas models \cite{RitortSollich} are simple caricatures of glassformers which
 employ local steric constraints as their sole means to glassiness
in the absence of any non-trivial static correlations
between particles.  It has been shown that, despite their apparent
simplicity, these models exhibit surprisingly many of the hallmarks of
glassy behavior that have been the focus of recent experimental and
theoretical efforts \cite{KobAndersenKLG, RitortSollich, Toninelli1, PanGarrahanChandler1,
  PanGarrahanChandler2}.  The model we study 
consists of hard core
particles on a triangular lattice with no static interactions other
than those that prohibit multiple occupancy of a single site.  To each
particle is associated a vector which can point along the six
directions bisecting the triangular lattice.  In other words, the
vectors point toward the interstitial sites of the lattice \cite{TLGrot}.  

Translation of particles obeys the kinetic constraints of the two
vacancy assisted triangular lattice gas, or the (2)-TLG \cite{Jackle1,
  Jackle2}: a particle at site
${\bf r}$ is
allowed
to move to a nearest neighbor site, ${\bf r}'$, if (1) ${\bf r}'$ is not occupied
and (2) the two mutual nearest neighbor sites of ${\bf r}$ and 
${\bf r}'$ are also empty.  These rules coincide with a physical
interpretation of steric constraints on the movement of hard core particles in a dense
fluid \cite{Jackle1}.

Rotation of particles obeys a similarly physically motivated kinetic
constraint if one imagines that the particles have small hard
protrusions along the direction of their orientation vector \cite{TLGrot}: a
particle, $i$, with a unit vector $\mathbf{p}_i$, can rotate either 60 degrees
clockwise or counterclockwise provided the two neighboring lattice
sites along the direction between the initial and final orientations
of $\mathbf{p}_i$ are empty.  Translations preserve the direction of
$\mathbf{p}_i$.  We refer to this model as the rotational TLG \cite{TLGrot}.  Due to the 
absence of non-trivial static interactions, there are
no static correlations between particles.  However, constraints on the
kinetics allow non-trivial dynamic correlations to emerge in
trajectory space.

In the computer
simulations, we investigated particle
densities, $\rho$, between 0.10 and 0.81 on a lattice with edge length
$L$ = 128 ($L$ = 256 for $\rho$ = 0.81).   The density $\rho$ = 1 corresponds to the
completely full lattice.  At densities up to and including 0.77, over 60
independent trajectories of lengths 10-100
times $\tau_{\alpha}$, where $\tau_{\alpha}$ is the time for the self-intermediate
  scattering
function at ${\bf q} = (\pi, 0)$ \cite{qnote} to reach $1/e$ of its
initial value (see below), were run.  At densities 0.79, 0.80 and
0.81, four to sixteen
trajectories were run.  These trajectories were stored
logarithmically for later analysis (i.e. configurations were saved
after 1, 2, 4, 8, 16, 32, etc. sweeps).  Time was measured in Monte
Carlo sweeps.  
During each sweep, particles
were chosen randomly and translational and orientational moves were
attempted with equal probability.  For the higher
density runs, a continuous time algorithm was used for
greater efficiency \cite{NewmanandBarkema}.  This algorithm involved making and updating a
list of only those particles which have the possibility of either
moving or rotating and
choosing from among those exclusively during every move.  The time was
then incremented as one over the number of possible moves.  
Due to the lack of static correlations, initial configurations
could be generated by random occupation of empty lattice sites by
particles with random orientations until the
desired density was reached.

\section{Heterogeneous dynamics and rotational correlation times}
\label{hetero}

Despite the lack of static correlations between rotational and translational degrees of freedom, 
dynamic coupling exists via the kinetic constraints \cite{TLGrot}.  A particle which cannot rotate because it is 
sterically blocked by neighboring particles must wait for those particles to translate away before it is
allowed to rotate again.  In this sense, we expect rotational dynamics to be a good indicator of local
structural relaxation in this model, as they are in experiment. 

Fig.\ \ref{probes} shows rotational trajectories of six particles at low and high densities.  The
particles at low density (Fig.\ \ref{probes}a) rotate freely, performing a random walk through all angles.
At higher densities, the particles perform random walks punctuated by periods of little to no
rotation.  In contrast to the trajectories at low density, dynamics at high density are clearly heterogeneous:
at any given time, some particles are rotating quickly while others
are essentially frozen.  Similar behavior has been observed for
translational motion
of a probe molecule immersed in spin-facilitated models \cite{Fickian,
  Jung1} and in poilymer films \cite{VandenboutScience, Vandenbout2001}.  

\begin{figure}
\centering
\includegraphics[width = 5in]{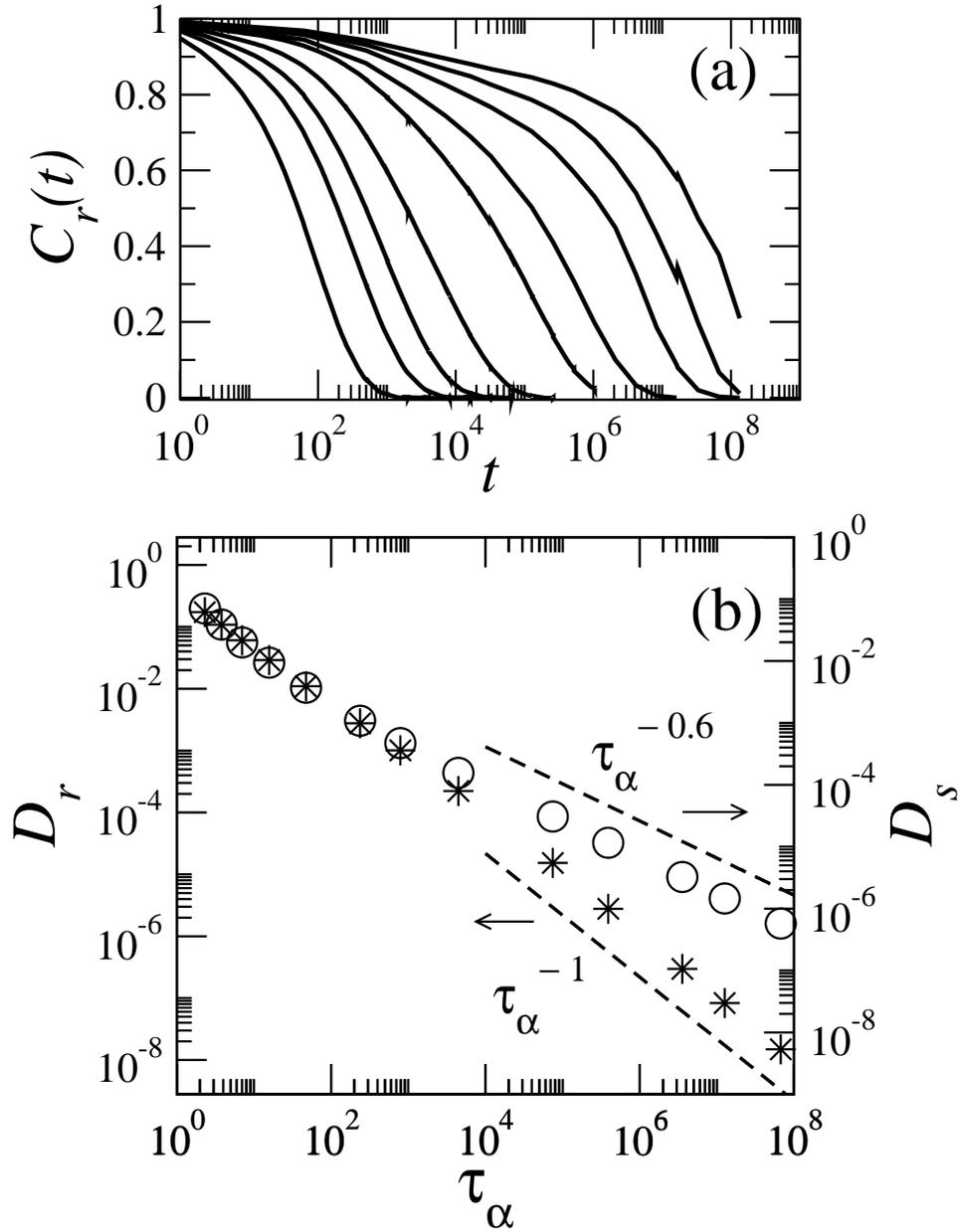}
\caption{(a) Rotational autocorrelation at (from left to right) $\rho$
= 0.50, 0.60, 0.65, 0.70, 0.75, 0.77, 0.79 and 0.81.  (b) Scaling of the rotational diffusion
constant, $D_r$ (stars), and translational self-diffusion constant,
$D_s$ (open circles), 
with structural relaxation time, $\tau_{\alpha}$.  The dashed lines
are the power laws, $\tau_{\alpha}^{-1}$ and $\tau_{\alpha}^{-0.6}$ as
indicated.}
\label{corr}
\end{figure}

A useful ensemble measure of slow dynamics is the rotational autocorrelation function, 
$C_r(t) = \langle {\mathbf{p}_i(0)\cdot \mathbf{p}_i(t)} \rangle$, which indicates
the time it takes for
a particle to lose memory of its initial spin orientation
\cite{VandenboutScience, TLGrot}.  Here, the angled brackets denote an average over
all particles and times, $t$.  A plot of $C_r(t)$ is shown in Fig.\ \ref{corr}a.  At low densities, relaxation
shows a simple exponential profile.  As density increases, the curves become more and more stretched
exponential indicative of averaging over multiple relaxation timescales.   This stretched
exponential decay is what one would expect qualitatively 
from rotational trajectories such as those depicted in Fig. \ref{probes}b.  An important quantity
that can be extracted from $C_r(t)$ is the rotational correlation time,
$\tau_r$, which is defined as $C_r(\tau_r) = 1/e$.  The inverse of
this timescale is the rotational diffusion constant, $D_r = \tau_r^{-1}$.  
 
Translational relaxation is often studied via the self-intermediate
scattering function, $F_s(q, t)$ = $\la e^{i{\bf q}\cdot({\bf r}_i(t) - {\bf r}_i(0))}\ra$.
Here, ${\bf r}_i(t)$ denotes the position of particle $i$ at time
$t$.  The decay of
the scattering function to $1/e$ at wavevector $\mathbf{q} = (\pi, 0)$ is typically
defined to be a structural relaxation time, $\tau_{\alpha}$, as
it
gives a sense of how density fluctuations relax at relatively short
lengthscales.  Dynamic behavior at large lengthscales is studied via
the self-diffusion constant, $D_s$, extracted from 
the mean-squared displacement, $\la|\Delta {\bf r}_i(t)|^2 \ra$ = $\la|{\bf
  r}_i(t) - {\bf r}_i(0)|^2\ra$.  The self-diffusion coefficient, $D_s$, is
defined as $D_s = \lim_{t\rightarrow\infty}\la|\Delta {\bf r}_i(t)|^2\ra/4t$.  
We omit further discussion of these quantities as they
have been presented at length elsewhere for this
\cite{PanGarrahanChandler1, Jackle1} and other models \cite{KobAndersenKLG}.

An important ramification of heterogeneous
dynamics is the breakdown of mean-field dynamical relations such
as the much studied Stokes-Einstein (SE) relation \cite{Sillescu1994,
  SwallenEdiger, Jung1}.  
In a system with homogeneous dynamics such as a normal liquid, we expect
relaxation behavior to be similar at all but the smallest lengthscales.  In a glass,
the presence of dynamic heterogeneity implies that such mean-field relations can be
violated.  In Fig. \ref{corr}b, we plot the rotational diffusion constant, $D_r$, and the
self-diffusion constant, $D_s$, versus the translational structural relaxation time, 
$\tau_{\alpha}$.  Rotational diffusion tracks structural relaxation whereas self-diffusion
does not.  That is, $D_r \sim \tau_{\alpha}^{-1}$ whereas $D_s \sim \tau_{\alpha}^{0.6}$.  
The same trend has been seen in experiment \cite{Sillescu1994}.  The
scaling of $D_r$ with $\tau_{\alpha}$ can be rationalized qualitatively
from the idea mentioned above that the ability to rotate is intimately tied to local structure.  

\begin{figure}
\centering
\includegraphics*[width=5in]{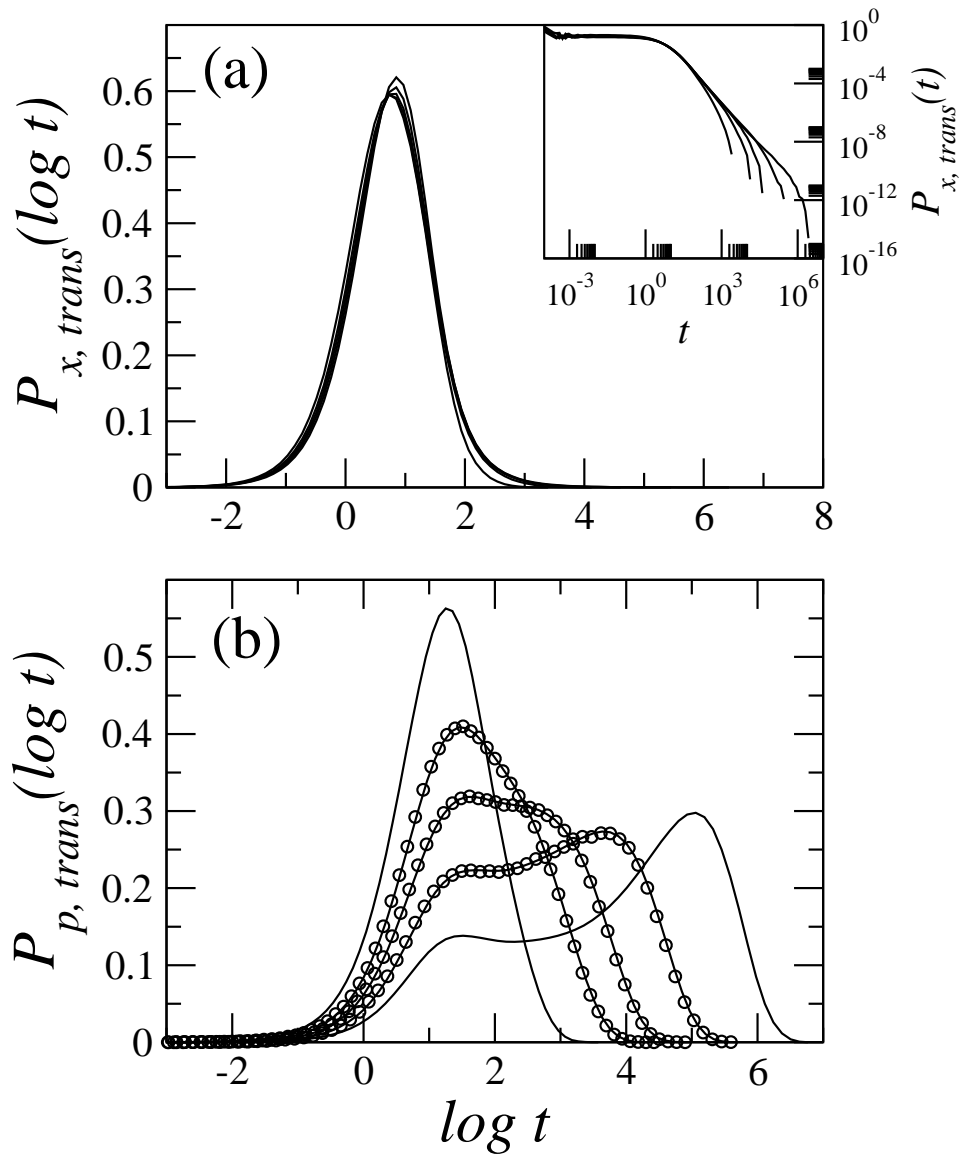}
\caption{Distributions of translational (a) exchange and (b)
  persistence times at (from left to right) $\rho$
= 0.50, 0.60, 0.65, 0.70 and 0.75.  The inset to (a) shows the exchange time
distributions as a function of linear time, $t$, as opposed to
  logarithmic time to emphasize the emergence of broader long time
  tails as density is increased.  The two distributions are related
  via
$P_{x, trans}(\log t) = t P_{x, trans}(t) \ln 10$.  The solid lines in
  (b) are the distributions of persistence
  times
calculated from the distributions of exchange times via eqn.\
  \ref{jungeq} 
and the open circles are results of direct calculations.} 
\label{trans}
\end{figure}

\section{Distributions of exchange and persistence times}
\label{dist}
 
\begin{figure}
\centering
\includegraphics*[width=5in]{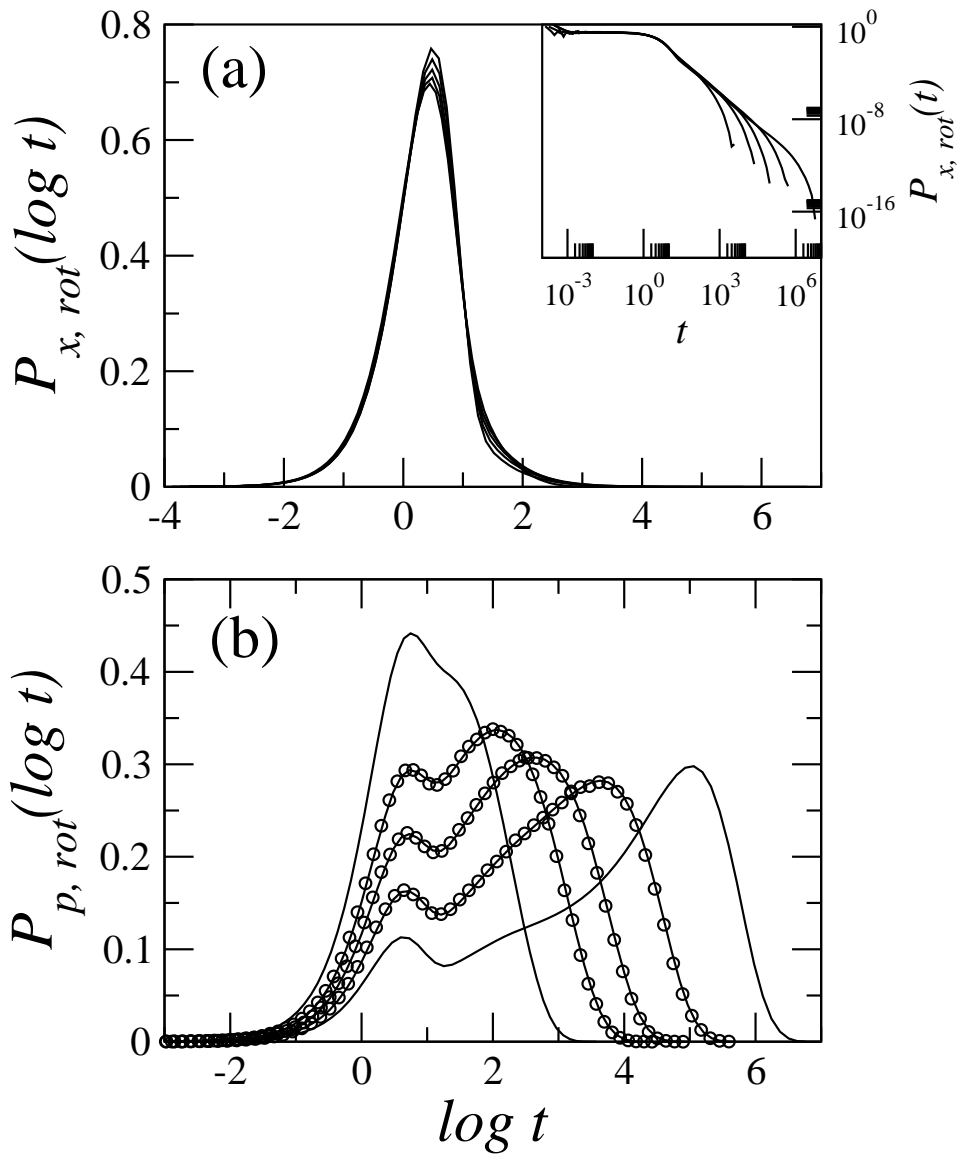}
\caption{Same as Fig. \ref{trans}, except for rotational times.} 
\label{rot}
\end{figure}

   Direct measures of heterogeneous dynamics in glassy
systems are distribution functions of persistence, $P_p(t)$, and
exchange, $P_x(t)$,
times \cite{VandenboutScience, GarrahanBerthier, Jung1, Jung2}.  Persistence times measure the 
first instance of a change in state
  given
an initial configuration. 
Exchange times measure the
duration of particular states.   
   For example, the distribution of rotational
persistence times, $P_{p, rot}(t)$, in this model is the distribution of times, given an
initial
configuration, when a particle changes its rotational
state for the first time.  The distribution of rotational exchange 
times, $P_{x, rot}(t)$, is the length of time a particle remains in a particular 
rotational state.  These distributions
  are multi-point functions because they depend not only on two points in
  time, but on all intervening points as well.  Fig. \ref{trans}a and
  Fig. \ref{rot}a show the distribution of exchange times for
  translational and rotational motion, respectively.  Distributions of 
exchange times are important for understanding the origin of dynamical
decoupling
phenomenon \cite{Jung1, Jung2} (see Fig.\ \ref{corr}b). 

The
distribution of exchange times for translations and rotations are very
similar.  There is very little change in short time structure as
$\rho$ increases.  Evidence of broadly distributed dynamics occurs in
the long time tails.  The distributions of persistence times for
translation
and rotation are also qualitatively similar and display the same
features noted in previous studies \cite{GarrahanBerthier,
  Jung1, PanGarrahanChandler1}.  Differences in structure between
rotational and translational persistence can be
attributed to the differing kinetic constraints.  
A particle can have the ability to rotate while being
constrained translationally and vice versa.

Recently, it has been shown that the distributions of exchange and persistence times
are related via the equation \cite{Jung2}:
\begin{equation}
\label{jungeq}
P_p(t) \sim \int_t^{\infty} P_x(t') dt'.
\end{equation}
The constant of
proportionality is fixed by normalization.  
This should be a general relation independent of the model or
observable studied.  The solid lines in Fig. \ref{trans}b and
  Fig. \ref{rot}b show the distribution of persistence times
  calculated via equation (\ref{jungeq}).  The open circles are the
  results of direct calculations.  We see that equation (\ref{jungeq}) is well-satisfied.

\section{Comparison with experiments}
\label{exp}

Recent single molecule experiments determined distributions of
rotational exchange times by following the dipole of embedded dye
molecules \cite{VandenboutScience,
  Vandenbout2001, Vandenbout2002}.  
Near the glass transition temperature, 
the absolute value of the rate of angular change, $\left |
\Delta\theta / \Delta t \right |$, 
showed abrupt changes
between different dynamical environments.  We can anticipate from
Fig. \ref{probes}b
that the same quantity in the rotational TLG will show similar
behavior.  From this quantity, Deschenes and Vandenbout extracted a 
distribution of exchange times using a standard deviation criterion \cite{Vandenbout2001}.
Here, we examine to what extent the distribution of times
measured in this way corresponds to the distribution of rotational
exchange times defined in section \ref{dist}.  

\begin{figure*}
\centering
\includegraphics[width=6in]{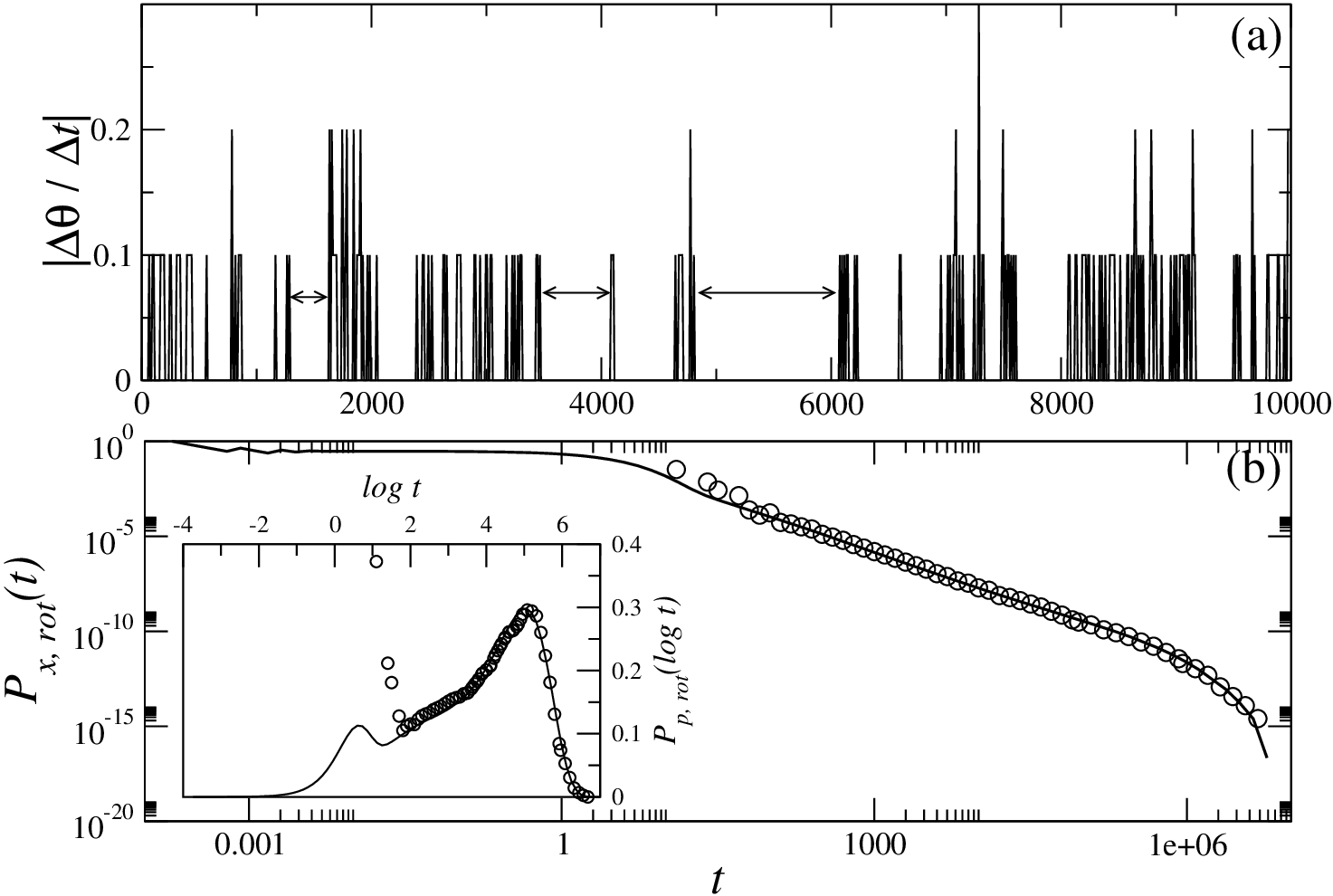}
\caption{(a) A plot of the absolute rate of angular change, 
$\left | \Delta\theta / \Delta t \right |$, for
  a rotational trajectory of a single particle at high density ($\rho$
  = 0.75).  Three examples of exchange times are indicated by the
  double headed arrows.  (b) Comparison
  of distributions of rotational exchange times obtained via data like
  those shown in panel (a) (open circles) and
  directly (solid line, see Fig.\ \ref{rot}a, inset).  The inset shows
  a comparison of the distribution
  of rotational persistence times.  The values for the open circles
  in the inset are obtained by applying eqn.\ \ref{jungeq} to the the data in
  the main part of panel (b).} 
\label{experiment}
\end{figure*}

In Fig. \ref{experiment}a, we plot $\left | \Delta\theta / \Delta t \right |$
where $\Delta \theta = \theta (t + \Delta t) - \theta (t)$ for the
rotational
TLG at high density.  Deschenes and Vandenbout assigned
an exchange event whenever the average angle jump changed by more than
two standard deviations from the previous average angle jump.  Due to
the coarse grained nature of the rotational TLG, $\left | \Delta\theta
/ \Delta t \right |$ changes by discrete jumps.  Therefore, exchange
events can be unambiguously assigned whenever such a jump occurs.  
Fig. \ref{experiment}a plots $\left | \Delta \theta / \Delta t \right |$ in
increments of ten sweeps during a portion of a single molecule
trajectory.  
The fastest exchange event
measurable at this time resolution is 10 sweeps.  
With infinite time resolution, however, we see that the 
definition of exchange times used in section \ref{dist} would
correspond precisely with the Deschenes-Vandenbout procedure. 

We verify this in Fig. \ref{experiment}b.  Here, the distribution
of times obtained via the Deschenes-Vandenbout procedure outlined
above (open circles) \cite{opencirclesamp}  are compared to
the distribution 
of rotational exchange times calculated
in section \ref{dist} (solid line).  The comparison is very good for
about ten orders of magnitude.  This method overestimates the
distribution at early times for the reasons of time resolution
mentioned in the previous paragraph.  The inset shows a comparison of
the distribution of rotational persistence times.  The open circles
are obtained from the open circles in the main panel of
Fig. \ref{experiment}b via equation (\ref{jungeq}).  Once again, the
agreement is very good.  In particular, the data obtained via
the procedure of Deschenes and Vandenbout captures the
structure and location of the main peak.

Single molecule experiments also measured the mean rotational exchange
times, $\tau_{x, rot}$ (i.e. the first moment of $P_{x, rot}(t)$,
FIG.\ \ref{rot}a, inset),
as a function of temperature.  It was found that $\tau_{x, rot}$
scaled with temperature in the same way as $\tau_r$, the rotational
correlation time, and that $\tau_{x, rot}$ was 10-20 times larger
than $\tau_r$ at a given temperature \cite{Vandenbout2001,
  Vandenbout2002}.  Our calculations indicate that 
$\tau_{x, rot} \sim (1-\rho)^{-2}$ or, using the definition that
$1-\rho \equiv \exp (-1/T)$ \cite{PanGarrahanChandler1}, 
$\tau_{x, rot} \sim \exp (-2/T)$.  The rotational correlation time,
$\tau_r$, scales like the structural relaxation time: $\ln
\tau_{\alpha} \sim \exp(1.7/T)$ \cite{PanGarrahanChandler1, Toninelli1}.
That is, $\tau_{x, rot}$ 
exhibits a much weaker dependence on temperature than 
$\tau_{r}$ similar to what is found in spin-facilitated models
\cite{Jung2}.  We also find that $\tau_{x, rot}$ is as much as six
orders of magnitude
times smaller than $\tau_r$ at the highest densities.  This
difference
with experimental findings is most likely due to the issue of time
resolution mentioned earlier and also discussed in
\cite{Vandenbout2002} which biases experimental measures of $\tau_{x,
  rot}$ to longer times. 
    
\acknowledgments 

The author would like to acknowledge David Chandler, Juan Garrahan and
Younjoon Jung for valuable discussions and the US National Science Foundation
and Department of Energy (grant no.\ DE-FE-FG03-87ER13793) for funding. 
The author was an NSF graduate research fellow for part of this work.


\begin{thebibliography}{21}
\bibitem[{qno()}]{qnote}
\bibinfo{note}{The wavevector dependent quantity, $F_s(q, t)$, was conveniently
  calculated on the square lattice because a correspondence can be made between
  the reciprocal lattice vectors of the square and triangular lattices as in
  \cite{Jackle1}.}
\bibitem[{ope()}]{opencirclesamp}
\bibinfo{note}{The data shown by the open circles have been shifted vertically
  to facilitate comparison with the exact distribution of exchange times.}
\end{thebibliography}
\end{document}